\documentclass[%
 reprint,
 amsmath,amssymb,
 aps,
 prc,
 superscriptaddress,
 longbibliography,
]{revtex4-2}

\usepackage{graphicx}% Include figure files
\usepackage{dcolumn}% Align table columns on decimal point
\usepackage{bm}% bold math
\usepackage[colorlinks,breaklinks]{hyperref}% add hypertext capabilities

\begin{document}

\title{
Exact construction and uniqueness of the coupled-channel Green's function
}

\author{Hao Liu}
\affiliation{School of Physics Science and Engineering, Tongji University, Shanghai 200092, China}
\author{Jin Lei}
\email[Contact author: ]{jinl@tongji.edu.cn}
\affiliation{School of Physics Science and Engineering, Tongji University, Shanghai 200092, China}
\affiliation{Southern Center for Nuclear-Science Theory (SCNT), Institute of Modern Physics,
 Chinese Academy of Sciences, Huizhou 516000, Guangdong Province, China}
\author{Zhongzhou Ren}
\affiliation{School of Physics Science and Engineering, Tongji University, Shanghai 200092, China}

\date{\today}

\begin{abstract}
We present a rigorous construction and uniqueness proof of the matrix Green's function for coupled radial Schr\"{o}dinger equations with symmetric coupling potentials. The Green's matrix $\mathbf{G}(R,R')$ is built from two fundamental sets of $N$ linearly independent solutions, regular and outgoing, of the coupled radial equations. We prove that the associated Wronskian matrix is diagonal with elements $W_n = -k_n$ and independent of the radial coordinate, and demonstrate through the symplectic structure of the $2N$-dimensional phase space that the resulting construction is the unique Green's matrix satisfying the defining equation with correct boundary conditions, continuity at the source point, and the prescribed derivative discontinuity. The construction applies to any system of coupled radial Schr\"{o}dinger equations with symmetric coupling potentials and open channels, including coupled-channels problems arising in nuclear, atomic, and molecular scattering. As an illustrative application, we show how the Green's matrix enters the nonlocal dynamical polarization potential within the continuum-discretized coupled-channels framework, where retaining the off-diagonal elements captures multistep excitation pathways beyond the weak-coupling approximation.\end{abstract}

\maketitle

\section{Introduction}

The Green's function of a system of coupled radial Schr\"{o}dinger equations is a fundamental object in quantum scattering theory. It governs the propagation of a particle or system through a set of coupled channels and appears naturally whenever one reduces a many-channel problem to an effective single-channel description. In nuclear, atomic, and molecular physics, coupled-channel Green's functions arise in the construction of optical potentials, the evaluation of transition amplitudes, and the treatment of resonance phenomena. Despite their central role, the construction of the full matrix Green's function for $N$ coupled channels, and the proof that the standard construction is the unique solution satisfying the correct boundary conditions, has received surprisingly little attention in the literature.

For a single uncoupled channel, the radial Green's function is a textbook result: it is built from two linearly independent solutions (regular and irregular) of the homogeneous equation, joined at the source point with the appropriate derivative discontinuity, and normalized by the Wronskian of the two solutions. The extension to $N$ coupled channels requires $2N$ linearly independent solution vectors organized into two $N \times N$ fundamental solution matrices, and the scalar Wronskian is replaced by a Wronskian matrix. This multichannel construction was developed in the context of atomic and molecular physics by Levine and Soven~\cite{PhysRevA.29.625} and further discussed by Levine~\cite{PhysRevA.30.1120}. Those references \emph{constructed} a coupled-channel Green's function by first building the regular and irregular solutions through eigenchannel ($K$-matrix) decomposition, and then \emph{verified} that the resulting bilinear formula satisfies the correct Wronskian relations and boundary conditions. The subsequent Brief Report by Levine~\cite{PhysRevA.30.1120} addressed the important practical problem of numerical stabilization during integration. The Levine-Soven analysis establishes that the bilinear formula is \emph{a} Green's function. It does not address whether it is the only one, and the structural properties of the Wronskian matrix and the reciprocity of the Green's matrix are imposed and verified rather than derived from the defining equation itself. In the present work we close this gap: starting from the defining equation and the regular/outgoing boundary conditions, we derive the bilinear formula directly and prove uniqueness through the symplectic structure of the $2N$-dimensional phase space of the coupled equations. The constancy and diagonal form of the Wronskian matrix, and the reciprocity of the Green's matrix under the exchange $(R,\gamma)\leftrightarrow(R',\gamma')$, then emerge as necessary consequences of the construction rather than as properties to be checked case by case. An analogous connection between transfer matrices and matrix Green's functions for systems of coupled second-order differential equations has been developed within the surface Green function matching (SGFM) framework by Rodriguez-Coppola \textit{et al.}~\cite{RodriguezCoppola90} and treated systematically in the monograph by P\'erez-\'Alvarez and Garc\'ia-Moliner~\cite{PerezAlvarezGarciaMoliner2004}, where the symplectic character of the transfer matrix and its relationship to the matrix Green function are central themes. That framework, however, addresses boundary-value matching problems in slab geometries rather than the radial scattering problem with regular and outgoing-wave boundary conditions relevant for coupled-channel reactions.

The coupled-channel Green's function finds important applications across many areas of physics. Wherever the Feshbach projection formalism~\cite{Feshbach58,Feshbach62} is used to reduce a multichannel problem to an effective single-channel description, the Green's function of the eliminated channels serves as the propagator that generates the nonlocal, energy-dependent polarization potential~\cite{Nagarajan85,Satchler91,Thompson89}. A particularly demanding application is the scattering of weakly bound nuclei~\cite{CANTO20061,CANTO20151}, where the continuum-discretized coupled-channels (CDCC) method~\cite{Rawitscher74,Kamimura86,AUSTERN1987125,10.1093/ptep/pts008} yields large systems of coupled equations with strong continuum-continuum couplings. The resulting dynamical polarization potential (DPP) has been shown to produce long-range attraction and absorption for halo projectiles~\cite{Mackintosh04,Keeley09}, and its microscopic construction has acquired renewed importance at rare-isotope beam facilities~\cite{Hebborn_2023}. Most previous studies have adopted a weak-coupling approximation that neglects the continuum-continuum couplings and renders the Green's matrix diagonal~\cite{Rao1973,PhysRevC.77.034609,Potel25}, but recent work has shown that the off-diagonal elements generate coherent interference contributions to the total absorption~\cite{Liu2025CoherentFusion}. We emphasize that this simplification is specific to the Green's-matrix construction of the DPP and is not a feature of loosely-bound-nucleus reaction theory in general. The continuum-discretized coupled-channels method itself retains the full continuum-continuum coupling, as do the adiabatic approximation for deuteron-induced reactions~\cite{JohnsonSoper70} and its halo-nucleus extension~\cite{JohnsonAlKhaliliTostevin97} through closure on the projectile internal Hamiltonian, the Weinberg state expansion that systematically improves the latter~\cite{JohnsonTandy74,LaidTostevinJohnson93,PangTimofeyukJohnsonTostevin13}, and the Faddeev treatment of three-body reaction dynamics~\cite{HlopheLei19,LazauskasCarbonell11,DeltuvaMoro07}. The weak-coupling simplification does not enter the solution of the three-body dynamics; it is introduced only when that dynamics is reduced to an effective two-body interaction through a diagonal Green's matrix. The present work removes this approximation by constructing the exact coupled-channel Green's matrix for the finite CDCC model space, retaining all continuum-continuum couplings among the included bins together with the exact channel-energy structure, from which the effective two-body interaction follows; this full coupled-channel Green's function is precisely the object whose construction and uniqueness we establish in this work. Analogous coupled-channel Green's functions appear in electron-atom and electron-molecule collisions~\cite{PhysRevA.29.625}, multichannel ultracold atomic scattering with multiple hyperfine channels, and coupled vibrational-rotational dynamics.

The central results of this paper are as follows. Starting from the defining equation and boundary conditions, we derive the coupled-channel Green's matrix $\mathbf{G}(R,R')$ and prove that the bilinear construction in terms of regular and outgoing solution matrices is the unique solution. We show that the associated Wronskian matrix is necessarily diagonal with elements $W_n = -k_n$ and independent of the radial coordinate, provided that the coupling potential matrix is symmetric, and that the symmetry of the Green's matrix under channel and coordinate exchange is a consequence of the symplectic structure of the $2N$-dimensional phase space. The vanishing of the self-Wronskians of the regular and irregular solution matrices plays a central role in this proof by eliminating cross terms in the matching conditions. The construction and uniqueness proof apply to any system of $N$ coupled radial Schr\"{o}dinger equations with symmetric coupling potentials and open channels, regardless of the physical origin of the coupling. As an illustrative application, we show how the Green's matrix enters the nonlocal DPP within the CDCC framework, retaining all continuum-continuum couplings. We also discuss the numerical challenges of inward propagation of irregular solutions and the stabilization techniques required to maintain accuracy.

The numerical implementation of this formalism and its application to deuteron-induced reactions on $^{58}$Ni are presented in a companion paper~\cite{Liu2025ExactDPP}, where the full-coupling effective potential is shown to reproduce CDCC observables exactly, while the weak-coupling approximation exhibits significant deviations.

The remainder of this paper is organized as follows. Section~II presents the construction of the coupled-channel Green's function from general fundamental solution matrices, proves the Wronskian properties, and establishes uniqueness through the symplectic structure of the $2N$-dimensional phase space. Section~III illustrates the formalism by deriving the nonlocal dynamical polarization potential within the CDCC framework. Section~IV provides a summary and outlook.

\section{Formalism}

We consider a general system of $N$ coupled radial Schr\"{o}dinger equations with symmetric coupling potentials. The Green's matrix elements $g_{\gamma\gamma'}(R,R')$ for this system satisfy the defining equation
\begin{equation}
\sum_{\beta}\mathcal{L}_{\gamma\beta}(R)\,g_{\beta\gamma'}(R,R')=\delta_{\gamma\gamma'}\,\delta(R-R'),
\label{eq.Gdef}
\end{equation}
where the differential operator is
\begin{equation}
\mathcal{L}_{\gamma\beta}(R) = \left(E_\gamma - T_\gamma\right)\delta_{\gamma\beta} - V_{\gamma\beta}(R),
\label{eq.Lop}
\end{equation}
with
\begin{equation}
T_\gamma = -\frac{\hbar^2}{2\mu}\left(\frac{d^2}{dR^2} - \frac{L_\gamma(L_\gamma+1)}{R^2}\right)
\end{equation}
being the radial kinetic energy operator for channel $\gamma$, $\mu$ the reduced mass, $E_\gamma$ the channel energy, $L_\gamma$ the orbital angular momentum quantum number, and $V_{\gamma\beta}(R)$ the coupling potential matrix element connecting channels $\gamma$ and $\beta$. The coupled radial equations for the homogeneous problem read
\begin{equation}
(E_\gamma - T_\gamma) f_\gamma(R) - \sum_{\gamma'} V_{\gamma\gamma'}(R)\, f_{\gamma'}(R) = 0.
\label{eq.cc}
\end{equation}
For a system with $N$ coupled channels, this equation admits $2N$ linearly independent solutions, which can be organized into two fundamental sets: $N$ regular solutions and $N$ irregular (outgoing) solutions.

The $N$ regular solutions are defined by their behavior at small radial distance. Each regular solution vector $u^n$ ($n = 1, \ldots, N$) satisfies the boundary condition $u_\gamma^n(R \to 0) \propto R^{L_\gamma + 1}\,\delta_{\gamma n}$, meaning that at the origin, only the $n$-th channel component is nonzero, and it vanishes as $R^{L_\gamma+1}$ in accordance with the centrifugal barrier. The $N$ irregular solutions are defined by their asymptotic behavior at large distance. Each irregular solution vector $h^n$ satisfies $h_\gamma^n(R \to \infty) \to \delta_{\gamma n}\,H^{(+)}_{L_\gamma}(k_\gamma R)$ in open channels, where $H^{(+)}_{L_\gamma}$ is the outgoing Coulomb-Hankel function (or Riccati-Hankel function in the absence of the Coulomb interaction) and $k_\gamma = \sqrt{2\mu E_\gamma}/\hbar$ is the channel wave number. These irregular solutions are obtained by inward integration from a large matching radius, which poses specific numerical challenges that will be discussed below. These $2N$ solution vectors can be organized into two $N \times N$ solution matrices
\begin{equation}
\begin{aligned}
\mathbf{U} &= \begin{bmatrix} u^1 & \cdots & u^N \end{bmatrix} = \begin{bmatrix}
u_1^1 & \cdots & u_1^N \\
\vdots & \ddots & \vdots \\
u_N^1 & \cdots & u_N^N
\end{bmatrix}, \\
\mathbf{H} &= \begin{bmatrix} h^1 & \cdots & h^N \end{bmatrix} = \begin{bmatrix}
h_1^1 & \cdots & h_1^N \\
\vdots & \ddots & \vdots \\
h_N^1 & \cdots & h_N^N
\end{bmatrix},
\end{aligned}
\label{eq.UH}
\end{equation}
where the rows are labeled by the channel index $\gamma$ and the columns by the solution index $n$.

We will show below that the coupled-channel Green's function is uniquely determined by the defining equation and boundary conditions. Collecting the elements $g_{\gamma\gamma'}(R,R')$ into the $N\times N$ Green's matrix $\mathbf{G}(R,R')$, with $[\mathbf{G}(R,R')]_{\gamma\gamma'}=g_{\gamma\gamma'}(R,R')$, we will show that it takes the form (cf.\ Refs.~\cite{PhysRevA.29.625,PhysRevA.30.1120})
\begin{equation}
\mathbf{G}(R,R') = \frac{2\mu}{\hbar^2} \left\{\begin{aligned}
\mathbf{U}(R)\, \mathbf{W}^{-1}\, \mathbf{H}^T(R') & \quad R < R', \\
\mathbf{H}(R)\, \mathbf{W}^{-1}\, \mathbf{U}^T(R') & \quad R > R',
\end{aligned}\right.
\label{eq.g_full}
\end{equation}
where $\mathbf{W}$ is the Wronskian matrix defined by
\begin{equation}
\mathbf{W}(R) = \mathbf{U}^T(R)\, \frac{d}{dR}\mathbf{H}(R) - \left[\frac{d}{dR}\mathbf{U}^T(R)\right] \mathbf{H}(R).
\label{eq.Wdef}
\end{equation}
The structure of this expression merits careful explanation. For $R < R'$, the Green's function is built from the regular solutions at the smaller coordinate and the irregular solutions at the larger coordinate, ensuring that $\mathbf{G}(R,R')$ satisfies the correct boundary conditions as $R \to 0$ and $R' \to \infty$. The Wronskian matrix $\mathbf{W}$ encodes the linear independence of the two fundamental sets and plays the role of a normalization factor. Its properties, constancy and diagonal structure, are crucial for the practical utility of Eq.~\eqref{eq.g_full} and are established in what follows. In the remainder of this section we first establish the properties of $\mathbf{W}$ taking the form~\eqref{eq.g_full} as given, and then prove, starting from the defining equation~\eqref{eq.Gdef}, that this form is its unique consequence.

We now demonstrate that, when the coupling potential matrix is symmetric ($V_{\gamma\gamma'} = V_{\gamma'\gamma}$), the Wronskian matrix $\mathbf{W}$ is independent of the radial coordinate $R$ and is, moreover, a diagonal matrix. The element $W_{nm}$ of the Wronskian matrix is given explicitly by
\begin{equation}
W_{nm} = \sum_{\gamma=1}^N \left( u_\gamma^n\, {h_\gamma^{m}}' - {u_\gamma^{n}}'\, h_\gamma^m \right),
\label{eq.Wnm}
\end{equation}
where primes denote derivatives with respect to $R$. Taking the derivative of $W_{nm}$ with respect to $R$ yields
\begin{equation}
\frac{d}{dR} W_{nm} = \sum_{\gamma=1}^N \left( u_\gamma^n\, {h_\gamma^{m}}'' - {u_\gamma^{n}}''\, h_\gamma^m \right).
\label{eq.dWnm}
\end{equation}
The second derivatives appearing here can be expressed through the coupled-channel equations~\eqref{eq.cc}. After multiplying by $2\mu/\hbar^2$ and rearranging, any solution $f_\gamma$ of the coupled equations satisfies
\begin{equation}
\begin{aligned}
{f_\gamma}'' &= \sum_{\gamma'}\mathcal{M}_{\gamma\gamma'}(R)\,f_{\gamma'},\\
\mathcal{M}_{\gamma\gamma'}(R) &= \left[\frac{L_\gamma(L_\gamma+1)}{R^2}-\frac{2\mu}{\hbar^2}E_\gamma\right]\delta_{\gamma\gamma'} + \frac{2\mu}{\hbar^2}V_{\gamma\gamma'}(R).
\end{aligned}
\label{eq.M}
\end{equation}
Substituting the second-derivative expressions for $u_\gamma^n$ and $h_\gamma^m$ into Eq.~\eqref{eq.dWnm}, we obtain
\begin{equation}
\frac{d}{dR} W_{nm}
= \sum_{\gamma=1}^N\sum_{\gamma'=1}^N
\left(u_\gamma^n\,\mathcal{M}_{\gamma\gamma'}\,h_{\gamma'}^m - h_\gamma^m\,\mathcal{M}_{\gamma\gamma'}\,u_{\gamma'}^n\right).
\label{eq.dWnm2}
\end{equation}
The symmetry of the coupling potential, $V_{\gamma\gamma'} = V_{\gamma'\gamma}$, implies that $\mathcal{M}_{\gamma\gamma'} = \mathcal{M}_{\gamma'\gamma}$. Consequently, the two terms in the double sum are related by the interchange $\gamma \leftrightarrow \gamma'$, and the entire expression vanishes identically: $dW_{nm}/dR = 0$. This establishes that each element of the Wronskian matrix is a constant, independent of the radial coordinate $R$, and its value is therefore determined once and for all by the boundary conditions.

The constant values of $W_{nm}$ are most conveniently evaluated from the asymptotic forms of the solutions. In the limit $R \to \infty$, the regular and irregular solutions approach
\begin{equation}
\begin{aligned}
u_\gamma^n &\rightarrow \frac{i}{2} \left( H^{(-)}_{L_\gamma}(k_\gamma R)\, \delta_{n\gamma} - S_{n\gamma}\, H^{(+)}_{L_\gamma}(k_\gamma R) \right), \\
h_\gamma^n &\rightarrow \delta_{n\gamma}\, H^{(+)}_{L_\gamma}(k_\gamma R),
\end{aligned}
\label{eq.asympt}
\end{equation}
where $H^{(\pm)}_{L}$ are the incoming and outgoing Coulomb-Hankel functions and $S_{n\gamma}$ are the elements of the scattering $S$ matrix. For the off-diagonal elements with $n \neq m$, the asymptotic boundary condition $h_\gamma^m \to \delta_{\gamma m}\,H^{(+)}_{L_\gamma}(k_\gamma R)$ ensures that only the $\gamma = m$ term survives in the sum over channels. However, this surviving term involves the Wronskian of $H^{(+)}$ with itself, $H^{(+)}{H^{(+)}}' - {H^{(+)}}'H^{(+)} = 0$, which vanishes identically. Therefore $W_{nm} = 0$ for all $n \neq m$, and the Wronskian matrix is diagonal.

For the diagonal elements ($n = m$), only the $\gamma = n$ term contributes, again because $h_\gamma^n$ vanishes asymptotically for $\gamma \neq n$. Evaluating the remaining term using Eq.~\eqref{eq.asympt} gives
\begin{equation}
\begin{aligned}
W_n &= u_n^n\, {h_n^{n}}' - {u_n^{n}}'\, h_n^n \\
&= \frac{i}{2} \Big[ \left( H^{(-)}_{L_n}(k_n R) - S_{nn}\, H^{(+)}_{L_n}(k_n R) \right) {H^{(+)}_{L_n}}'(k_n R) \\
&\quad - \left( {H^{(-)}_{L_n}}'(k_n R) - S_{nn}\, {H^{(+)}_{L_n}}'(k_n R) \right) H^{(+)}_{L_n}(k_n R) \Big] \\
&= \frac{i}{2} \left[ H^{(-)}_{L_n}(k_n R)\, {H^{(+)}_{L_n}}'(k_n R) - {H^{(-)}_{L_n}}'(k_n R)\, H^{(+)}_{L_n}(k_n R) \right] \\
&= -k_n,
\end{aligned}
\label{eq.Wn}
\end{equation}
where the terms proportional to $S_{nn}$ cancel, and the final equality follows from the standard Wronskian relation of the Coulomb-Hankel functions under the Riccati normalization convention adopted throughout this work. Different conventions for the Hankel functions modify only the overall constant prefactor and sign of $W_n$, without affecting the physical content of the Green's function. Thus the Wronskian matrix takes the simple diagonal form $\mathbf{W} = \text{diag}(-k_1, -k_2, \ldots, -k_N)$.

With the diagonal Wronskian established, the Green's function of Eq.~\eqref{eq.g_full} can be written in a form that makes the role of each linearly independent solution explicit. Expanding the matrix products, the individual matrix elements become
\begin{equation}
g_{\gamma\gamma'}(R, R') = \frac{2\mu}{\hbar^2} \left\{ \begin{aligned}
\sum_n \frac{u_\gamma^n(R)\, h_{\gamma'}^n(R')}{W_n} & \quad R < R', \\
\sum_n \frac{h_\gamma^n(R)\, u_{\gamma'}^n(R')}{W_n} & \quad R > R',
\end{aligned}\right.
\label{eq.g_elements}
\end{equation}
where the sum runs over the $N$ solution indices. Each term in this sum represents the contribution of one linearly independent solution pair to the propagation from channel $\gamma'$ at $R'$ to channel $\gamma$ at $R$. For symmetric couplings, the Green's matrix also satisfies the reciprocity relation $g_{\gamma\gamma'}(R,R') = g_{\gamma'\gamma}(R',R)$, which provides a useful numerical consistency check during implementation.

We now derive the Green's matrix of Eq.~\eqref{eq.g_full} directly from the defining equation~\eqref{eq.Gdef} and the boundary conditions, and prove that it is the unique solution. Unlike the original construction~\cite{PhysRevA.29.625,PhysRevA.30.1120}, which arrived at the bilinear formula through an eigenchannel-based procedure and then verified its correctness, the derivation below obtains the formula as the only possible outcome of the boundary conditions and matching requirements. As a byproduct, the Wronskian constancy $W_n = -k_n$ and the reciprocity of the Green's matrix emerge as necessary consequences, rather than properties that need to be checked separately. The proof proceeds in three steps: establishing the self-Wronskian identities, demonstrating the uniqueness of the functional form, and solving the matching conditions at the source point.

For any two $N \times N$ solution matrices $\mathbf{X}$ and $\mathbf{Y}$ of the coupled radial equations~\eqref{eq.cc}, the bilinear concomitant $\mathcal{W}[\mathbf{X},\mathbf{Y}] \equiv \mathbf{X}^T\mathbf{Y}' - \mathbf{X}'^T\mathbf{Y}$ is independent of $R$ when $V_{\gamma\gamma'} = V_{\gamma'\gamma}$, by the same argument used above for $\mathbf{W} = \mathcal{W}[\mathbf{U},\mathbf{H}]$. Applying this result to the self-Wronskians of the fundamental solution matrices, we find that for the regular solutions, the small-$R$ boundary condition $u_\gamma^n(R \to 0) \propto R^{L_\gamma+1}\delta_{\gamma n}$ implies that every element of $\mathbf{U}^T\mathbf{U}' - \mathbf{U}'^T\mathbf{U}$ vanishes at the origin, since the product of any two regular solution components vanishes as $R \to 0$. By constancy, this gives
\begin{equation}
\mathcal{W}[\mathbf{U},\mathbf{U}] = 0.
\label{eq.WUU}
\end{equation}
Similarly, the large-$R$ boundary condition $h_\gamma^n(R \to \infty) \to \delta_{\gamma n}H^{(+)}_{L_\gamma}(k_\gamma R)$, together with the vanishing of the single-function Wronskian $H^{(+)}{H^{(+)}}' - {H^{(+)}}'H^{(+)} = 0$ and the column-wise orthogonality of the asymptotic conditions, yields
\begin{equation}
\mathcal{W}[\mathbf{H},\mathbf{H}] = 0.
\label{eq.WHH}
\end{equation}

The Green's matrix $\mathbf{G}(R,R')$ must be regular as $R \to 0$ (to avoid unphysical singularities at the origin) and purely outgoing as $R \to \infty$ (to enforce causality). For a fixed source point $R'$, the most general matrix function with these properties can be expressed as
\begin{equation}
\mathbf{G}(R,R') = \begin{cases}
\mathbf{U}(R)\,\mathbf{A}(R'), & R < R',\\
\mathbf{H}(R)\,\mathbf{B}(R'), & R > R',
\end{cases}
\label{eq.Gform}
\end{equation}
where $\mathbf{A}$ and $\mathbf{B}$ are $N \times N$ coefficient matrices to be determined. This is the only form compatible with the boundary conditions, since any regular solution of the homogeneous equation is a linear combination of the columns of $\mathbf{U}$, and any outgoing solution is a linear combination of the columns of $\mathbf{H}$. No ansatz is involved in writing Eq.~\eqref{eq.Gform}; it follows uniquely from the completeness of the fundamental solution sets.

The coefficient matrices are determined by the matching conditions at the source point $R = R'$. Continuity of the Green's function requires
\begin{equation}
\mathbf{U}(R')\,\mathbf{A}(R') = \mathbf{H}(R')\,\mathbf{B}(R'),
\label{eq.cont}
\end{equation}
while integrating the defining equation~\eqref{eq.Gdef} across the delta-function source at $R = R'$ yields the derivative jump condition
\begin{equation}
\mathbf{H}'(R')\,\mathbf{B}(R') - \mathbf{U}'(R')\,\mathbf{A}(R') = \frac{2\mu}{\hbar^2}\,\mathbf{I}_N.
\label{eq.jump}
\end{equation}
We now solve for $\mathbf{B}$. Left multiplying the continuity condition~\eqref{eq.cont} by $\mathbf{U}'^T$ gives
\begin{equation}
\mathbf{U}'^T \mathbf{U}\,\mathbf{A} = \mathbf{U}'^T \mathbf{H}\,\mathbf{B},
\label{eq.step1}
\end{equation}
while left multiplying the jump condition~\eqref{eq.jump} by $\mathbf{U}^T$ gives
\begin{equation}
\mathbf{U}^T \mathbf{H}'\,\mathbf{B} - \mathbf{U}^T \mathbf{U}'\,\mathbf{A} = \frac{2\mu}{\hbar^2}\,\mathbf{U}^T.
\label{eq.step2}
\end{equation}
Subtracting Eq.~\eqref{eq.step1} from Eq.~\eqref{eq.step2}, we collect the terms containing $\mathbf{A}$ and $\mathbf{B}$ separately:
\begin{equation}
\begin{aligned}
&\underbrace{(\mathbf{U}^T \mathbf{H}' - \mathbf{U}'^T \mathbf{H})}_{\displaystyle =\,\mathbf{W}}\,\mathbf{B}\\
&\quad -\;\underbrace{(\mathbf{U}^T \mathbf{U}' - \mathbf{U}'^T \mathbf{U})}_{\displaystyle =\,\mathcal{W}[\mathbf{U},\mathbf{U}]\,=\,0}\,\mathbf{A}
= \frac{2\mu}{\hbar^2}\,\mathbf{U}^T.
\end{aligned}
\end{equation}
The first bracket is precisely the Wronskian matrix $\mathbf{W}$ as defined in Eq.~\eqref{eq.Wdef}. The second bracket is the self-Wronskian $\mathcal{W}[\mathbf{U},\mathbf{U}]$, which vanishes identically by Eq.~\eqref{eq.WUU}. This eliminates $\mathbf{A}$ entirely, leaving
\begin{equation}
\mathbf{W}\,\mathbf{B}(R') = \frac{2\mu}{\hbar^2}\,\mathbf{U}^T(R'),
\end{equation}
from which
\begin{equation}
\mathbf{B}(R') = \frac{2\mu}{\hbar^2}\,\mathbf{W}^{-1}\,\mathbf{U}^T(R').
\label{eq.B}
\end{equation}

The derivation of $\mathbf{A}$ follows the same strategy with a different choice of multipliers. Left multiplying Eq.~\eqref{eq.cont} by $\mathbf{H}'^T$ gives
\begin{equation}
\mathbf{H}'^T \mathbf{U}\,\mathbf{A} = \mathbf{H}'^T \mathbf{H}\,\mathbf{B},
\label{eq.step3}
\end{equation}
and left multiplying Eq.~\eqref{eq.jump} by $\mathbf{H}^T$ gives
\begin{equation}
\mathbf{H}^T \mathbf{H}'\,\mathbf{B} - \mathbf{H}^T \mathbf{U}'\,\mathbf{A} = \frac{2\mu}{\hbar^2}\,\mathbf{H}^T.
\label{eq.step4}
\end{equation}
Subtracting Eq.~\eqref{eq.step3} from Eq.~\eqref{eq.step4}:
\begin{equation}
\begin{aligned}
&\underbrace{(\mathbf{H}^T \mathbf{H}' - \mathbf{H}'^T \mathbf{H})}_{\displaystyle =\,\mathcal{W}[\mathbf{H},\mathbf{H}]\,=\,0}\,\mathbf{B}\\
&\quad -\;\underbrace{(\mathbf{H}^T \mathbf{U}' - \mathbf{H}'^T \mathbf{U})}_{\displaystyle =\,-\mathbf{W}^T}\,\mathbf{A}
= \frac{2\mu}{\hbar^2}\,\mathbf{H}^T.
\end{aligned}
\end{equation}
The first bracket is the self-Wronskian $\mathcal{W}[\mathbf{H},\mathbf{H}] = 0$ by Eq.~\eqref{eq.WHH}, which eliminates $\mathbf{B}$. For the second bracket, we note that $\mathbf{H}^T\mathbf{U}' - \mathbf{H}'^T\mathbf{U}$ is the negative transpose of the Wronskian: since $\mathbf{W} = \mathbf{U}^T\mathbf{H}' - \mathbf{U}'^T\mathbf{H}$, taking the transpose gives $\mathbf{W}^T = \mathbf{H}'^T\mathbf{U} - \mathbf{H}^T\mathbf{U}'$, so that $\mathbf{H}^T\mathbf{U}' - \mathbf{H}'^T\mathbf{U} = -\mathbf{W}^T$. The equation therefore reduces to
\begin{equation}
\mathbf{W}^T\,\mathbf{A}(R') = \frac{2\mu}{\hbar^2}\,\mathbf{H}^T(R'),
\end{equation}
from which
\begin{equation}
\mathbf{A}(R') = \frac{2\mu}{\hbar^2}\,\mathbf{W}^{-T}\,\mathbf{H}^T(R').
\label{eq.A}
\end{equation}
Since $\mathbf{W}$ is diagonal for symmetric couplings, we have $\mathbf{W}^{-T} = \mathbf{W}^{-1}$. Substituting the expressions~\eqref{eq.B} and~\eqref{eq.A} for $\mathbf{B}$ and $\mathbf{A}$ back into Eq.~\eqref{eq.Gform} reproduces exactly the construction of Eq.~\eqref{eq.g_full}.

It is instructive to verify that the resulting Green's matrix is indeed self-consistent by checking that the continuity condition~\eqref{eq.cont} and the jump condition~\eqref{eq.jump} are both satisfied. Substituting Eqs.~\eqref{eq.B} and~\eqref{eq.A} into the continuity condition~\eqref{eq.cont}, we need $\mathbf{U}\,\mathbf{A} = \mathbf{H}\,\mathbf{B}$, which after inserting the explicit expressions becomes
\begin{equation}
\mathbf{U}\,\mathbf{W}^{-1}\,\mathbf{H}^T = \mathbf{H}\,\mathbf{W}^{-1}\,\mathbf{U}^T,
\label{eq.symmetry}
\end{equation}
where the argument $R'$ and the common factor $\frac{2\mu}{\hbar^2}$ have been suppressed for brevity. This identity states that the matrix $\mathbf{C}\equiv\mathbf{U}\mathbf{W}^{-1}\mathbf{H}^T$ is equal to its own transpose $\mathbf{C}^T=\mathbf{H}\mathbf{W}^{-1}\mathbf{U}^T$. To see why this is nontrivial, note that the element $(\gamma,\gamma')$ of $\mathbf{C}$ and $\mathbf{C}^T$ reads
\begin{equation}
C_{\gamma\gamma'}=\sum_n\frac{u_\gamma^n\,h_{\gamma'}^n}{W_n},\qquad
C^T_{\gamma\gamma'}=C_{\gamma'\gamma}=\sum_n\frac{u_{\gamma'}^n\,h_\gamma^n}{W_n}.
\end{equation}
These two sums are manifestly different term by term: for each $n$, the outer product $u^n h^{nT}$ (with elements $u_\gamma^n h_{\gamma'}^n$) is not the same matrix as $h^n u^{nT}$ (with elements $h_\gamma^n u_{\gamma'}^n$), yet the claim is that after summing over all $n$ with weights $1/W_n$, the result is symmetric. This is a completeness property of the full set of $2N$ fundamental solutions, and we now prove it using the symplectic structure of the solution space.

We assemble the $2N$ fundamental solutions and their derivatives into the $2N\times 2N$ phase-space matrix
\begin{equation}
\boldsymbol{\Phi}(R) = \begin{pmatrix} \mathbf{U}(R) & \mathbf{H}(R) \\ \mathbf{U}'(R) & \mathbf{H}'(R) \end{pmatrix},
\label{eq.Phi}
\end{equation}
whose upper-left and upper-right $N\times N$ blocks contain the solution values, and whose lower blocks contain their derivatives. The key observation is that the coupled radial equations preserve a symplectic structure encoded in the $2N\times 2N$ matrix $\mathbf{J}=\bigl(\begin{smallmatrix}0 & \mathbf{I}_N \\ -\mathbf{I}_N & 0\end{smallmatrix}\bigr)$. The generalized Wronskian $\boldsymbol{\Phi}^T\mathbf{J}\boldsymbol{\Phi}$ is constant in $R$ (as a consequence of the symmetry $V_{\gamma\gamma'}=V_{\gamma'\gamma}$, by the same argument that proved constancy of $\mathbf{W}$). Computing this product block by block,
\begin{equation}
\begin{aligned}
\boldsymbol{\Phi}^T \mathbf{J}\,\boldsymbol{\Phi}
&= \begin{pmatrix}
\mathbf{U}^T\mathbf{U}'-\mathbf{U}'^T\mathbf{U} & \mathbf{U}^T\mathbf{H}'-\mathbf{U}'^T\mathbf{H} \\
\mathbf{H}^T\mathbf{U}'-\mathbf{H}'^T\mathbf{U} & \mathbf{H}^T\mathbf{H}'-\mathbf{H}'^T\mathbf{H}
\end{pmatrix}\\
&= \begin{pmatrix}
\mathcal{W}[\mathbf{U},\mathbf{U}] & \mathbf{W} \\
-\mathbf{W}^T & \mathcal{W}[\mathbf{H},\mathbf{H}]
\end{pmatrix},
\end{aligned}
\label{eq.PhiJPhi}
\end{equation}
where we have identified the four blocks: the $(1,1)$ block is $\mathcal{W}[\mathbf{U},\mathbf{U}]=0$ by Eq.~\eqref{eq.WUU}, the $(1,2)$ block is $\mathbf{W}$ by its definition~\eqref{eq.Wdef}, the $(2,2)$ block is $\mathcal{W}[\mathbf{H},\mathbf{H}]=0$ by Eq.~\eqref{eq.WHH}, and the $(2,1)$ block is $\mathbf{H}^T\mathbf{U}'-\mathbf{H}'^T\mathbf{U}=-\mathbf{W}^T=-\mathbf{W}$ (the last equality holds because $\mathbf{W}$ is diagonal). Thus
\begin{equation}
\boldsymbol{\Phi}^T \mathbf{J}\,\boldsymbol{\Phi}
= \begin{pmatrix}
0 & \mathbf{W} \\
-\mathbf{W} & 0
\end{pmatrix}
\equiv \boldsymbol{\Omega}.
\label{eq.Omega}
\end{equation}
Since $W_n=-k_n\neq 0$ for all channels, $\mathbf{W}$ is invertible, and so is $\boldsymbol{\Omega}$ with inverse
\begin{equation}
\boldsymbol{\Omega}^{-1}
= \begin{pmatrix}
0 & -\mathbf{W}^{-1} \\
\mathbf{W}^{-1} & 0
\end{pmatrix},
\label{eq.Omega_inv}
\end{equation}
as can be verified directly from $\boldsymbol{\Omega}\,\boldsymbol{\Omega}^{-1}=\mathbf{I}_{2N}$. Since $\det\boldsymbol{\Omega}\neq 0$ and $\boldsymbol{\Phi}^T\mathbf{J}\boldsymbol{\Phi}=\boldsymbol{\Omega}$, the matrix $\boldsymbol{\Phi}$ itself must be invertible. From $\boldsymbol{\Phi}^T\mathbf{J}\boldsymbol{\Phi}=\boldsymbol{\Omega}$ we obtain $\boldsymbol{\Phi}^T\mathbf{J}=\boldsymbol{\Omega}\,\boldsymbol{\Phi}^{-1}$, and therefore
\begin{equation}
\boldsymbol{\Phi}^{-1} = \boldsymbol{\Omega}^{-1}\,\boldsymbol{\Phi}^T\,\mathbf{J}.
\label{eq.Phiinv_formula}
\end{equation}
We now evaluate the right-hand side step by step. First,
\begin{equation}
\boldsymbol{\Phi}^T\,\mathbf{J}
= \begin{pmatrix}
\mathbf{U}^T & \mathbf{U}'^T \\
\mathbf{H}^T & \mathbf{H}'^T
\end{pmatrix}
\begin{pmatrix}
0 & \mathbf{I}_N \\
-\mathbf{I}_N & 0
\end{pmatrix}
= \begin{pmatrix}
-\mathbf{U}'^T & \mathbf{U}^T \\
-\mathbf{H}'^T & \mathbf{H}^T
\end{pmatrix}.
\label{eq.PhiTJ}
\end{equation}
Then, left multiplying by $\boldsymbol{\Omega}^{-1}$,
\begin{equation}
\begin{aligned}
\boldsymbol{\Phi}^{-1}
&= \begin{pmatrix}
0 & -\mathbf{W}^{-1} \\
\mathbf{W}^{-1} & 0
\end{pmatrix}
\begin{pmatrix}
-\mathbf{U}'^T & \mathbf{U}^T \\
-\mathbf{H}'^T & \mathbf{H}^T
\end{pmatrix}\\
&= \begin{pmatrix}
\mathbf{W}^{-1}\mathbf{H}'^T & -\mathbf{W}^{-1}\mathbf{H}^T \\
-\mathbf{W}^{-1}\mathbf{U}'^T & \mathbf{W}^{-1}\mathbf{U}^T
\end{pmatrix}.
\end{aligned}
\label{eq.Phiinv}
\end{equation}
We can now extract the desired identities from the completeness relation $\boldsymbol{\Phi}\,\boldsymbol{\Phi}^{-1}=\mathbf{I}_{2N}$. Writing out the block matrix product explicitly,
\begin{equation}
\begin{pmatrix}
\mathbf{U} & \mathbf{H} \\
\mathbf{U}' & \mathbf{H}'
\end{pmatrix}
\begin{pmatrix}
\mathbf{W}^{-1}\mathbf{H}'^T & -\mathbf{W}^{-1}\mathbf{H}^T \\
-\mathbf{W}^{-1}\mathbf{U}'^T & \mathbf{W}^{-1}\mathbf{U}^T
\end{pmatrix}
=\begin{pmatrix}
\mathbf{I}_N & 0 \\
0 & \mathbf{I}_N
\end{pmatrix}.
\label{eq.completeness}
\end{equation}
Each of the four $N\times N$ blocks in this equation yields an independent identity. The upper-right $(1,2)$ block gives
\begin{equation}
\begin{aligned}
&\mathbf{U}\bigl(-\mathbf{W}^{-1}\mathbf{H}^T\bigr)
+\mathbf{H}\bigl(\mathbf{W}^{-1}\mathbf{U}^T\bigr)\\
&\quad= -\mathbf{U}\,\mathbf{W}^{-1}\,\mathbf{H}^T + \mathbf{H}\,\mathbf{W}^{-1}\,\mathbf{U}^T = 0,
\end{aligned}
\label{eq.cont_check}
\end{equation}
which is precisely the continuity condition~\eqref{eq.symmetry}. This identity is the completeness relation that guarantees $\mathbf{C}=\mathbf{C}^T$, despite the individual outer products being non-symmetric. The lower-right $(2,2)$ block gives
\begin{equation}
\begin{aligned}
&\mathbf{U}'\bigl(-\mathbf{W}^{-1}\mathbf{H}^T\bigr)
+\mathbf{H}'\bigl(\mathbf{W}^{-1}\mathbf{U}^T\bigr)\\
&\quad= \mathbf{H}'\,\mathbf{W}^{-1}\,\mathbf{U}^T - \mathbf{U}'\,\mathbf{W}^{-1}\,\mathbf{H}^T = \mathbf{I}_N.
\end{aligned}
\label{eq.jump_check}
\end{equation}
Substituting the explicit forms of $\mathbf{A}$ and $\mathbf{B}$ into the jump condition~\eqref{eq.jump}, the left-hand side reads
\begin{equation}
\begin{aligned}
\mathbf{H}'\mathbf{B}-\mathbf{U}'\mathbf{A}
&=\frac{2\mu}{\hbar^2}\left(\mathbf{H}'\mathbf{W}^{-1}\mathbf{U}^T-\mathbf{U}'\mathbf{W}^{-1}\mathbf{H}^T\right)\\
&=\frac{2\mu}{\hbar^2}\,\mathbf{I}_N,
\end{aligned}
\end{equation}
where the second equality uses Eq.~\eqref{eq.jump_check}. This matches the right-hand side of Eq.~\eqref{eq.jump} exactly. The remaining two blocks yield analogous identities involving derivatives: the upper-left $(1,1)$ block gives
\begin{equation}
\mathbf{U}\,\mathbf{W}^{-1}\,\mathbf{H}'^T - \mathbf{H}\,\mathbf{W}^{-1}\,\mathbf{U}'^T = \mathbf{I}_N,
\end{equation}
and the lower-left $(2,1)$ block gives
\begin{equation}
\mathbf{U}'\,\mathbf{W}^{-1}\,\mathbf{H}'^T - \mathbf{H}'\,\mathbf{W}^{-1}\,\mathbf{U}'^T = 0.
\end{equation}
All four blocks of the completeness relation are satisfied simultaneously, confirming that Eq.~\eqref{eq.g_full} is the unique Green's matrix satisfying the correct boundary conditions at the origin and at infinity, continuity at the source point, and the prescribed derivative discontinuity.

We emphasize that the construction and uniqueness proof presented above rely on three conditions: (i) the coupling potential matrix is symmetric, $V_{\gamma\gamma'} = V_{\gamma'\gamma}$; (ii) all $N$ channels are open, so that $k_\gamma > 0$ and the Wronskian matrix $\mathbf{W}$ is invertible; and (iii) the regular and outgoing boundary conditions are well defined. No assumption has been made about the physical origin of the coupled channels. The formalism therefore applies to any system of $N$ coupled radial Schr\"{o}dinger equations satisfying these conditions, including coupled-channels descriptions of inelastic nuclear scattering, electron-atom and electron-molecule collisions, ultracold atomic scattering with multiple hyperfine channels, and coupled vibrational-rotational dynamics. When closed channels are present ($k_\gamma^2 < 0$), the outgoing boundary condition is replaced by exponential decay, and the corresponding Wronskian elements acquire a modified form; the extension of the present proof to mixed open-closed systems requires a separate treatment of the closed-channel boundary conditions.

A practical issue that must be addressed in evaluating Eq.~\eqref{eq.g_elements} is the numerical stability of the irregular solutions during inward propagation. As discussed by Levine and Soven~\cite{PhysRevA.29.625} and Levine~\cite{PhysRevA.30.1120}, the irregular solution matrix $\mathbf{H}$ is obtained by integrating inward from the asymptotic region, where the column-wise boundary conditions are imposed. During this inward propagation, channels with different orbital angular momenta exhibit vastly different growth rates: in the approximately decoupled small-$R$ limit, the irregular solution in channel $\gamma$ behaves as $h_\gamma^n(R) \sim R^{-L_\gamma}$, while the regular solution scales as $u_\gamma^n(R) \sim R^{L_\gamma+1}$. As $L_\gamma$ increases, the dynamic range among channels grows rapidly, and the faster-growing components can numerically overwhelm the slower ones, leading to a loss of linear independence among the columns of $\mathbf{H}$. This contamination propagates into the off-diagonal elements of the Green's function and manifests as a drift of the computed Wronskian from its exact constant value. Reference~\cite{PhysRevA.30.1120} points out that such deviations can grow approximately exponentially if stabilization is not enforced during the propagation.

To maintain numerical accuracy, practical implementations should periodically re-stabilize or re-orthogonalize the irregular solution matrix during the inward integration, using techniques such as modified Gram-Schmidt orthogonalization or related procedures. The constancy of the Wronskian matrix provides a sensitive diagnostic: any deviation of the computed $W_n$ from $-k_n$ signals a loss of numerical precision. Even with careful stabilization, the Green's function construction encounters a fundamental limitation at sufficiently high partial waves, where the inward integration of irregular solutions becomes intrinsically ill conditioned. In applications, one should verify convergence with respect to $L_{\max}$ and, when necessary, treat very high-$L$ contributions with asymptotic or perturbative approximations.

\section{Application to breakup dynamical polarization potentials}

As an illustrative application, we show how the coupled-channel Green's function enters the construction of the dynamical polarization potential (DPP) within the continuum-discretized coupled-channels (CDCC) framework for the scattering of weakly bound composite projectiles. Weakly bound nuclei, such as $^{6}$He, $^{11}$Be, $^{8}$B, $^{6}$Li, $^{7}$Li, and $^{9}$Be, have separation energies of the order of one MeV or less, so that coupling between the ground state and the continuum profoundly modifies the reaction dynamics~\cite{CANTO20061,CANTO20151}. We consider a two-body projectile composed of a core $b$ and a valence particle $x$, incident on a target $A$; the internal projectile coordinate is denoted $\bm{r}$ and the projectile-target relative coordinate $R$. In the CDCC method~\cite{Rawitscher74,Kamimura86,AUSTERN1987125,10.1093/ptep/pts008}, the projectile continuum is discretized into $N$ square-integrable bin states $\phi_{bx}^\gamma$ ($\gamma = 1,\ldots,N$), built on the ground state $\phi_{bx}^0$, yielding a system of $N$ coupled radial equations of the form~\eqref{eq.cc} for the relative motion between the projectile and target. The channel energy is $E_\gamma = E - \epsilon_\gamma$, with $E$ the total center-of-mass energy and $\epsilon_\gamma$ the excitation energy of bin $\gamma$.

The Feshbach projection formalism~\cite{Feshbach58,Feshbach62} partitions the model space into the elastic ($P$) and continuum ($Q$) subspaces and yields an effective single-channel Hamiltonian for the elastic channel:
\begin{equation}
H_{\text{eff}} = PHP + PHQ \frac{1}{E + i\varepsilon - QHQ} QHP,
\label{eq.Heff}
\end{equation}
where $P = |\phi_{bx}^0\rangle\langle\phi_{bx}^0|$ projects onto the projectile ground state and $Q = \sum_{\gamma=1}^{N} |\phi_{bx}^{\gamma}\rangle\langle\phi_{bx}^{\gamma}|$ onto the discretized continuum. The first term contains the bare elastic-channel Hamiltonian, while the second term is the DPP, which is manifestly energy dependent and nonlocal. The DPP can be expressed in terms of the coupled-channel Green's matrix constructed in Sec.~II as
\begin{equation}
\Delta U(R,R') = \sum_{\gamma,\gamma'=1}^N U_{0\gamma}(R)\,g_{\gamma\gamma'}(R,R')\,U_{\gamma'0}(R'),
\label{eq.DPP}
\end{equation}
where $U_{0\gamma}(R) = \langle\phi_{bx}^0| U_{bA} + U_{xA} |\phi_{bx}^\gamma\rangle_{\bm{r}}$ are the coupling form factors between the ground state and the continuum bins, with $U_{bA}$ and $U_{xA}$ the core-target and valence-target optical potentials and the matrix element taken over the internal coordinate $\bm{r}$, and $g_{\gamma\gamma'}(R,R')$ are the matrix elements of the $Q$-space Green's operator in the channel basis. The physical interpretation is transparent: the projectile is excited from the ground state into a continuum bin $\gamma'$ at position $R'$ through the coupling $U_{\gamma'0}(R')$, propagates through the coupled continuum channels via $g_{\gamma\gamma'}(R,R')$, and is de-excited back to the ground state at position $R$ through $U_{0\gamma}(R)$. The sum over all intermediate channels encodes all possible excitation and de-excitation pathways. Applying the optical theorem within the $P$-space, the imaginary part of the expectation value of $\Delta U$ with respect to the elastic wave function provides a direct measure of the total reaction flux absorbed into all nonelastic channels.

Most previous studies~\cite{Rao1973,PhysRevC.77.034609,Potel25} have adopted a weak-coupling approximation in which the couplings among the continuum channels are neglected, reducing the Green's matrix to $g_{\gamma\gamma'} = g_\gamma \delta_{\gamma\gamma'}$. The DPP then simplifies to a sum of separable terms, $\Delta U(R,R') = \sum_\gamma U_{0\gamma}(R)\,g_\gamma(R,R')\,U_{\gamma 0}(R')$, which is computationally attractive but neglects the multistep propagation pathways through which continuum-continuum couplings redistribute flux among the breakup channels. Alternative strategies based on trivially equivalent local potentials~\cite{Canto09,rangel2025nucleusnucleuspotentialsscatteringtightly} suppress the intrinsic nonlocality and introduce partial-wave dependence that obscures the microscopic origin of the effective interaction. The full coupled-channel construction of Sec.~II retains all such pathways, including the coherent interference contributions highlighted in Ref.~\cite{Liu2025CoherentFusion}. The numerical implementation and application to deuteron scattering on $^{58}$Ni are presented in a companion paper~\cite{Liu2025ExactDPP}, where the full-coupling effective potential is shown to reproduce CDCC elastic scattering cross sections exactly, while the weak-coupling and folding-model approaches show significant deviations.

\section{Summary and Outlook}

In this work we have presented a rigorous construction and uniqueness proof of the matrix Green's function for $N$ coupled radial Schr\"{o}dinger equations with symmetric coupling potentials. The Green's matrix was built from two fundamental sets of solutions, regular and outgoing, organized into $N \times N$ solution matrices. While the bilinear form of the Green's matrix was previously obtained by Levine and Soven~\cite{PhysRevA.29.625,PhysRevA.30.1120} through an eigenchannel construction and verified by substitution, the present work derives it directly from the defining equation and boundary conditions and proves that it is the unique solution. The proof relies on the symplectic structure of the $2N$-dimensional phase space. The self-Wronskian identities $\mathcal{W}[\mathbf{U},\mathbf{U}] = 0$ and $\mathcal{W}[\mathbf{H},\mathbf{H}] = 0$ play a central role by eliminating the cross terms in the matching conditions, while the completeness relation $\boldsymbol{\Phi}\,\boldsymbol{\Phi}^{-1} = \mathbf{I}_{2N}$ guarantees the continuity and symmetry of the Green's matrix at the source point.

The construction and uniqueness proof hold for any system of coupled radial Schr\"{o}dinger equations with symmetric coupling potentials and open channels, regardless of the physical context. The same framework is directly applicable to coupled-channels descriptions of inelastic nuclear scattering, effective optical potential constructions via the Feshbach formalism, electron-atom and electron-molecule collisions, and multichannel ultracold scattering problems. Wherever the Feshbach projection formalism is used to reduce a multichannel problem to an effective single-channel description, the coupled-channel Green's function derived here provides the exact propagator for the eliminated channels.

As a specific application, we showed how the Green's matrix enters the nonlocal dynamical polarization potential within the CDCC framework, retaining all continuum-continuum couplings. The resulting DPP kernel, expressed as a bilinear sum over the full Green's matrix elements, naturally incorporates the off-diagonal propagation mechanisms through which continuum-continuum couplings generate additional nonlocality, energy dependence, and coherent interference in the elastic effective interaction.

We also discussed the practical challenges of implementing this construction, particularly the loss of numerical conditioning during inward propagation of irregular solutions at high partial waves. The constancy of the Wronskian provides a built-in diagnostic for monitoring numerical precision.

The numerical implementation of this framework, including its application to deuteron-induced reactions on $^{58}$Ni and a quantitative benchmarking of the weak-coupling approximation, is presented in a companion paper~\cite{Liu2025ExactDPP}. That work demonstrates that the full-coupling effective potential reproduces CDCC elastic scattering cross sections exactly across a range of incident energies, while the weak-coupling and folding-model approaches show significant deviations. Extensions to systematic studies of how breakup-induced nonlocality varies across different projectile-target systems and energy regimes are in progress.

\begin{acknowledgments}
This work was supported by the National Natural Science Foundation of China (Grants No.~12475132 and No.~12535009), the National Key R\&D Program of China (Contract No.~2023YFA1606503), and the Fundamental Research Funds for the Central Universities. During the preparation of this work, the authors used AI-based tools (large language model assistants) to support language editing and to cross-check the analytic derivations. All physics content and results were verified independently by the authors, who assume full responsibility for the work.
\end{acknowledgments}

\section*{Data availability}

There are no publicly available research data or software supporting this paper. Requests for further information or data should be sent to the authors.

\bibliography{cdcc}

\end{document}